\documentclass[onecolumn,11pt,  
 showpacs,  
 aps,  
 prd,  
 a4paper,  
 superscriptaddress,  
 nofootinbib,  
 tightenlines,  
 floats,floatfix  
 ]{revtex4}

\usepackage{graphicx}
\usepackage{epsfig}
\usepackage{subfigure}
\usepackage{latexsym}
\usepackage{amsmath,amssymb}        

\newcommand{\gsim}{\mbox{\raisebox{-1.ex}{$\stackrel
     {\textstyle>}{\textstyle\sim}$}}}
\newcommand{\lsim}{\mbox{\raisebox{-1.ex}{$\stackrel
     {\textstyle<}{\textstyle \sim}$}}}

\newcommand{\beq}{\begin{equation}}
\newcommand{\eeq}{\end{equation}}
\newcommand{\beqa}{\begin{eqnarray}}
\newcommand{\eeqa}{\end{eqnarray}}
\newcommand{\bea}{\begin{array}}
\newcommand{\ena}{\end{array}}

\input colordvi

\def\be{\begin{equation}}
\def\ee{\end{equation}}
\def\bea{\begin{eqnarray}}
\def\eea{\end{eqnarray}}

\def\4pig{\sfrac{4\pi G}{c^{4}}}

\def\hsp5{\hspace{5mm}}
\newcommand{\sfrac}[2]{{\textstyle{#1\over#2}}}
\def\case#1/#2{\textstyle\frac{#1}{#2}}

\begin{document}

\title{Inflation driven by q-de Sitter}

\author{M. R. Setare}
\email{rezakord@ipm.ir}
\affiliation{Department of Science, Campus of Bijar, University of Kurdistan, Bijar, Iran}

\author{D. Momeni}
\email{d.momeni@yahoo.com}
\affiliation{Eurasian International Center for Theoretical Physics and Department of General \& Theoretical Physics,
Eurasian National University, Astana 010008, Kazakhstan.
}
\author{V. Kamali}
\email{vkamali1362@gmail.com}
\affiliation{Department of Physics, Faculty of Science,
Bu-Ali Sina University, Hamedan, 65178, Iran}

\author{R. Myrzakulov}
\email{rmyrzakulov@gmail.com}
\affiliation{Eurasian International Center for Theoretical Physics and Department of General \&  Theoretical Physics,
Eurasian National University, Astana 010008, Kazakhstan.
}


\begin{abstract}
We propose a generalised de Sitter scale factor for the cosmology of early and late time universe, including single scalar field is called as inflaton. This form of scale factor has a free parameter $q$ is called as  nonextensivity parameter. When $q=1$, the scale factor is de Sitter. This scale factor is an intermediate form between power-law and de Sitter. We study cosmology of such families. We show that both kinds of dark components, dark energy and dark matter simultaneously are described by this family of solutions. As a motivated idea, we investigate inflation in the framework of $q$-de Sitter. We consider three types of scenarios for inflation. In a single inflation scenario, we observe that, inflation ended without any specific ending inflation $\phi_{end}$, the spectral index and the associated running of the spectral index are %
$ n_\mathrm{s} - 1 \sim -2\epsilon , \quad \alpha_\mathrm{s} \equiv 0 $. To end the inflation:
we should have $q=\frac{3}{4}$. We deduce that the inflation ends when the evolution of the scale factor is $a (t) =e_ {3/4} (t) $. With this scale factor there is no need to specify $\phi_{end}$. As an alternative to have inflation with ending point, We will study q-inflation model in the context of warm inflation. We propose two forms of  damping term $\Gamma$. In the first case when $\Gamma=\Gamma_0$, we show the scale invariant spectrum, (Harrison-Zeldovich spectrum, i.e. $n_s=1$) may be approximately presented by ($q=\frac{9}{10},~~N=70$). Also there is a range of values of $R$ and $n_s$ which is compatible with the BICEP2 data where $q=\frac{9}{10}$. In case $\Gamma=\Gamma_1V(\phi)$, it is observed that small values of a number of e-folds are assured for small values of $q$ parameter. Also in this case, the scale-invariant spectrum  may be represented by $(q,N)=(\frac{9}{10},70)$. For $q=\frac{9}{10}$ a range of values of $R$ and $n_s$ is compatible with the BICEP2 data. Consequently, the proposal of $q$-de Sitter is consistent with observational data. We observe that the non-extensivity parameter $q$ plays a significant role in inflationary scenario.

\end{abstract}

\maketitle


\section{Introduction}	
The inflationary idea was proposed to solve a list of cosmological problems which were appearing in the context of the standard
Big Bang theory \cite{Sta80,oldinf}, namely flatness problem\cite{WMAP,LLbook},horizon problem, the origin of large-scale structure in the universe and relic density problem. Now it is believed mostly that inflation is a viable and reasonable candidate describing the very early acceleration expansion of the Universe.One of the simplest candidate for inflation is a single scalar field in the framework of slow-roll approximation. This slow-role assumption is verified by different observational data like the
Cosmic Microwave Background (CMB) temperature
anisotropies measured by COBE \cite{COBE},
WMAP \cite{WMAP}, and Planck \cite{Planck}. From the perturbation point of view, curvature perturbations of the inflatons has a smooth scale-invariant
primordial power spectrum \cite{oldper}. Such basic property is fundamentally observed and is in great agreement  with
CMB anisotropies. The main problem concerning single inflation is how we realize or physically make it obvious . One way is to realize it as modifications of general relativity as modified gravities (see for review \cite{mog}). \par
From another point of view, we have more strong ideas about an accelerating expansion of the late time and existence of a type of phase transition of radiation to matter and dark energy, for example the data of the type I
Supernovae (SN Ia) \cite{Riess,Perlmutter} , CMB \cite{WMAP1} and
Baryon Acoustic Oscillations (BAO) \cite{BAO}. In the last phase, dark energy there are several proposals like a simple positive cosmological constant $\Lambda$ \cite{Weinberg,CST} which has a fine-tuning problem. Also a wide class of scalar fields proposed for dark energy \cite{quinpapers}.
(For a recent review on inflation in the framework of field theories see \cite{Tsujikawa:2014mba}.)
As we mentioned the modified gravities is more popular and as the first example it was the Starobinsky model in which we replace Einstein-Hilbert Lagrangian by a quadratic correction term $R\to f (R) =R+R^2/(6M^2) $ \cite{Sta80}, were here as GR, $R$ denotes Ricci scalar and $M$ is a mass parameter which we need to attend to the de Sitter space-time.
 We mention here that according to the recent observational constraints , the equation of state of dark energy which is the ratio between the effective pressure and dark energy density $w_{\rm DE}=P_{\rm DE}/\rho_{\rm DE} $ is in the interval of $w_{\rm DE} -1$ \cite{WMAP9,CDT,Planck}. One of the success of modified gravities of type f (R) is we able to explain $-1<w_{\rm DE} \leq -1$ without any need to have a ghost field~\cite{moreview} .
Warm inflation is also a way to attribute inflation which is recently motivated by authors \cite{5-w} in the context of general relativity and Tachyionic models to teleparallel  \cite{Jamil:2013nca}.

In this paper we study a class of cosmological solutions in which the scale factor evolves as a generalized de Sitter form. It is supposed that the time interval between initial and ending time of inflation is divided into infinite numbers of short time intervals in which in each of this interval de Sitter is the dominant evolution. But the general evolutionary scheme is a non additive function,is called $q$-exponential or $q$-de Sitter. This type of generalized functions firstly proposed in statistical mechanics for non extensive systems, thermodynamic systems far away from the equilibrium points. Entropy of many different physical systems is non-additive. If you merge two systems with different quantum states and a definite entropy form, the entropy of the merged system cannot be obtained by simple addition of two entropies. Indeed it is inspired from the second law of thermodynamics. In the black hole version, consider two black holes in thermodynamic equilibrium with entropies $S (1), S (2) $. If we merge these systems, the new black hole has new basic hairs, mass, charge and angular momentum. The final state has entropy $S (1+2) $ in which it always satisfies the following inequality, so called as the second law of thermodynamic:
\begin{eqnarray}
S(3)\geq S(1)+S(2).
\end{eqnarray}
The difference between the initial and final states is defined as the non extensivity , or generalized entropy form:
\begin{eqnarray}
S(1,2)=S(3)- S(1)-S(2).
\end{eqnarray}
It is straightforward to show that the common form of the entropy, the Boltzmann formula for distribution of quantum states is no longer valid for such systems.\par
So far the idea of generalized Boltzmann distributions or q-exponential functions have been motivated for different areas of physics, but not cosmology. In this paper we apply the idea of non extended to the time evolution of an early Universe.

In Sec.~\ref{q-exp} we review the mathematical theory of generalized exponential functions,or q-exponential functions. Our views are statistical physics and mathematical physics.

In Sec.\ref{cosmology} we investigate cosmological evolution of a flat  Friedmann-Lema\^{i}tre-Robertson-Walker (FLRW) universe. We show that how this type of scale factor unifies dark matter and dark energy in a same scenario.

In Sec.\ref{scalar} we study single inflaton models with q-exponential scale factor.

In Sec.\ref{warm}we investigate another scenario for inflation with a reheating mechanism.


Sec.~\ref{conclusion} is devoted to conclusions.
\par
Throughout the paper our units are adopted as the following,  $c=\hbar=1$ ,
 $c$ is the speed of light and $\hbar$
is Dirac constant, Planck mass
$M_{{\rm pl}}=2.4357\times10^{18}$\,GeV via
$\kappa^2\equiv 8\pi G=1/M_{{\rm pl}}^{2}$.
We adopt the metric signature $(-,+,+,+)$.

\section{$q$-Exponential functions:Quick review}\label{q-exp}
Let us  consider a bijective map with the following definition :
\begin{eqnarray}
f:=z\to e^{z},\ \ \mathcal{C}\to\mathcal{C}.
\end{eqnarray}
This function is called exponential and has the following basic properties:
\begin{eqnarray}
f(z+w)=f(z)f(w),\ \ f^{-1}(z)=\ln(z).
\end{eqnarray}
Physically it means our system has boosted or translational symmetry in a specific direction. The function has an inverse and one to one map each point of the domain $\mathcal{C}$ to a unique point on $\mathcal{C}$. We are dealing to interpret this exponential function as the physical probability of a distinct system formalized with a definite energy and temperature. In statistical physics, the canonical probability to find a system with the Hamiltonian $\mathcal{H}$ of a given system in temperature $T=\frac{1}{\beta}$ and in a definite energy $E$ in canonical ensemble is defined by:
\begin{eqnarray}
p_{\beta}(E)\sim e^{-\beta \mathcal{H}}.
\end{eqnarray}
The inverse of this function is defined entropy if we think on the $f (z) $ as the probability of a physical state in a distinct state $z$. The common form of entropy is :
\begin{eqnarray}
S\sim -\Sigma_{E_{i}}p_{\beta}(E_{i})\log \Big[p_{\beta}(E_{i})\Big]
\end{eqnarray}
This definition of entropy is additive.
For example a mixture of two systems $A, B$, the probability is $p_{\beta}(E_A+E_B) $, so the entropy of the numbers of the accession states of the combined system reads as $S (A+B) =S (A)+S (B) $. All these formulas derived from the basic assumption of the probability as the exponential function.\par
 But as we examine, in many important physical cases, especially for the black holes and for the cosmological backgrounds, the gravitational entropy of the horizon is not an additive function. It means we need to redefine the probability function. The appropriate form is not so long valid as the exponential form but in the other generalized form, in the texts known as the generalized exponential function. Inspired from the statistical physics, we defined the $q$-exponential function $e_ {q} (z) $ in the following form\cite{Tsallis}:
\begin{eqnarray}
e_{q}(z)=\Sigma_{n=0}^{\infty}\frac{z^n}{(q;q)_n}=\phi_{1;0}\Big[0|-1;q;z\Big].
\end{eqnarray}
Here the q-hypergeometric function is defined \cite{Exton}:
\begin{eqnarray}
&&\phi_{r;s}\Big[\alpha_1,\alpha_2,...,\alpha_r|\beta_1,\beta_2,...\beta_s;q;z\Big]=\Sigma_{n=0}^{\infty}\frac{\Pi_{i=1}^{r}(\alpha_i;q)_n}{\Pi_{i=1}^{s}(\beta_i;q)_n}
\times \frac{z^n}{(q;q)_n}\Big[(-1)^nq^{n(n-1)/2}\Big]^{1+s-r}.
\end{eqnarray}
Where $(a;q)_k$ is $q$-Pochhammer symbol as the following:
\begin{eqnarray}
&&(a;q)_k=\Pi_{j=0}^{k-1}(1-aq^j) \theta(k)+\delta_{k,1}+\Pi_{j=1}^{|k|}(1-aq^{-j})^{-1} \theta(-k)+\Pi_{j=0}^{\infty}(1-aq^j) \delta_{k,\infty}.
\end{eqnarray}
here $\theta(x)$ is Heaviside step function and $\delta_{i,j}$ is the Kronecker delta.
A more simplified form of this exponential function,which is more convenient for us is as the following:
\begin{eqnarray}
e_{q}(z)=\Big[1+(q-1)z\Big]^{\frac{1}{q-1}}.
\end{eqnarray}
Note that here $\lim_{q\to 1} e_{q}(z)=e^z$. It is the probability distribution function of a given physical system, if we minimize the generalized entropy $S_q(A)$  of the non-extensive systems:
\begin{eqnarray}
&&S_q({p_i}) = {1 \over q - 1} \left( 1 - \sum_i p_i^q \right),\ \ \sum_i p_i=1.
\end{eqnarray}
 For non-extensive systems the generalized entropy satisfies the following non-additivity condition(for a review see \cite{Tsallis (2009)}:
\begin{eqnarray}
&&S_{q}(A+B)=S_{q}(A)+S_{q}(B)+(1-q)S_{q}(A)S_{q}(B),\ \ k_{B}=1.
\end{eqnarray}
Let us to think on  $e^{Ht}$ as the scale factor of a de-Sitter Universe. Its exact solution of the cosmological FLRW equations with cosmological constant,and without any other form of matter fields:
\begin{eqnarray}
&&H=H_0,\ \ H\equiv \frac{\dot{a(t)}}{a(t)}.
\end{eqnarray}
Here $a(t) $ is scale factor of a flat FLRW Universe with the following metric:
\begin{eqnarray}
&&ds^2=-dt^2+a(t)^2d\vec{x}^2.
\end{eqnarray}
A solution of cosmological equations with $q$-exponential function $e_ {q} (Ht) $ is called as $q$-de Sitter. The $q$-modified FLRW equations are given by the following:
\begin{eqnarray}
&&3H_{q}(t)^2=\kappa^2\rho,\ \ 2(\frac{d}{dt})_qH_{q}(t)=-\kappa^2(p+\rho).
\end{eqnarray}
Here $(\frac{d}{dt})_qf(t)\equiv\frac{f(qt)-f(t)}{t(q-1)}$, $\lim_{q\to 1} (\frac{d}{dt})_qf(t)=\dot{f}(t)$ . We use the same symbol as $H_{q}(t)\equiv H(t)=(\frac{d}{dt})_q a (t) $. $q$-de Sitter is an exact solution for case $p=-\rho$.\par
The extended de-Sitter scale factor $a (t) =e_ {q} (Ht) $ has the limit of the commonly studied de-Sitter when we set $q=1$. But as we will study later, the cosmological implications of such $q$-de-Sitter Universe is widely different. We remember the following basic properties of the q-exponential function:
\begin{eqnarray}
&&e_{q}(z)e_{1/q}(z)=1,\ \ e_{q}(z)=E_q(z(1-q)),\\&&\nonumber
E_{q}(z)=\phi_{1,0}(0;q;z)=\Pi_{n=0}^{\infty}\Big(1-zq^n \Big)^{-1}.
\end{eqnarray}
We mention here that $e_ {q} (z) $ to be considered as the gene function of the q-derivative operator $(\frac{d}{dz})_q=D_q$ of a deformation of the commonly used derivatives:
\begin{eqnarray}
D_{q}e_{q}(z)=e_{q}(z).
\end{eqnarray}
In the following section we explore some cosmological aspects of $q$-de Sitter model.

\section{$q$-de-Sitter proposal for FLRW cosmology}\label{cosmology}
Our aim of this section is to investigate the cosmological behavior of $q$-de-Sitter family. We will replace this kind of scale factor in the common forms of FLRW equations \footnote{We assume that $q$-de-Sitter solves some kinds of cosmological equations with specific matter distributions. An alternative form is to rewrite all the gravitational equations and FLRW equations in terms of $q$-derivatives. This is a hard task and we omit it in our study. }.
From now to the ending on this article,we use the following representation of the $q$-de-Sitter scale factor for a flat FLRW metric:
\begin{eqnarray}
a(t)=e_{q}(H_0 t)=\Big[1+(q-1)H_0 t\Big]^{\frac{1}{q-1}}\label{a(t)}.
\end{eqnarray}
We observe that in the limit of $q\to 1$, this scale factor matches de Sitter form:
\begin{eqnarray}
a_{dS}(t)=\lim_{q\to 1} \Big[1+(q-1)H_0 t\Big]^{\frac{1}{q-1}}=e^{H_0 t}.
\end{eqnarray}
Furthermore, this type of scale factor interpolate between power-law and de Sitter,when we study its form near the early times, $H_0 t\gg 1$, we have:
\begin{eqnarray}
a_{early}(t)\sim \Big[H_0 t\Big]^{\frac{1}{q-1}}=t^{p}.
\end{eqnarray}
To have acceleration expansion, $p>1$, we should have $q<2$.  We conclude that:
\begin{eqnarray}
a_{early}(t)\preceq e_{q}(H_0 t)\preceq a_{dS}(t)\label{inequality}
\end{eqnarray}
Inequality which is presented in (\ref{inequality}) is one of the main reasons which we explore the cosmological evolution with $q$-de Sitter. It is a good candidate to connect smoothly early times to later times.

\par
To be more precise, let us review the cosmological behavior of the (\ref{a(t)}).
Think on a finite interval in the cosmological epoch, nearly the inflation with time interval $t\in[0,T]$. Consider a partionization of the interval to $[0,T]:=\cup_{i=0}^{N}[t_i,t_{i+1}], t_ {N+1} =T$.
At each interval $I_i= [t_i, t_ {i+1}] $ we assume that the scale factor is $q$-exponential $e_ {q} (H_0 (t_ {i+1} -t_ {i})) $. So, if we sum over all time intervals, we have a finite time $q$-exponential scale factor. We mention here that in each interval $I_i$ the cosmological evolution is not de-Sitter. But we have a $q$-de-Sitter solution. It is more interesting for the case in which we study inflation but without the de-Sitter scale factor. \par Let us to investigate the cosmology of a flat FLRW universe , filled with a perfect fluid with pressure $p$ and energy density $\rho$. The FLRW equations are written as the following:
\begin{eqnarray}\label{q}
&& 3H^2=\kappa^2 \rho,\\
&& 2\dot{H}=-\kappa^2(\rho+p).
\end{eqnarray}
Using the (\ref{a(t)}) give us:
\begin{eqnarray}\label{q}
&&H(t)=\frac{\dot{a}(t)}{a(t)}=\frac{H_0}{1+(q-1)H_0 t}=H_0\Big(e_{q}(H_0t)\Big)^{q-1}\label{H}
\end{eqnarray}
The effective equation of state parameter $w$ and deceleration parameter which we denote it here by $\tilde{q}$  read as the following:
\begin{eqnarray}
&&w=\frac{p}{\rho}=-\frac{1}{3}(1+2q)\label{w}\\&&
\nonumber
\tilde{q}=-1-\frac{\dot{H}}{H^2}=-q\label{q}.
\end{eqnarray}
All symbols have the same meaning as the other cosmological models. Just we used $\tilde{q}$ for the deceleration parameter to avoid of any confusion with the $q$ factor. We observe that for $q\to1$ we've the complete description of de-Sitter Universe:
\begin{eqnarray}
&&H(t)=H_0,\\&&
w=\frac{p}{\rho}=-1,\\&&
\tilde{q}=-1.
\end{eqnarray}
It means that late time (de-Sitter) is considered as a limiting (asymptotic $t\to \infty$) case. Our scale factor contains this late time epoch  in limiting case $q\to1$.
But for $q\neq1$, the fluid mimics, either dark energy neither dark matter at the same time:
\begin{eqnarray}
&&\texttt{Dark matter}:\ \ \Big(w=0\Big)	\equiv a(t)=e_{-1/2}(H_0 t),\\&&
\texttt{Dark energy}:\ \ \Big(-1<w<-\frac{1}{3}\Big)	\equiv a(t)=e_{0<q<1}(H_0 t).
\end{eqnarray}
To have a well defined scale factor for dark matter, we need to suppose that $|t|<\frac{2}{3H_0}$. As we know the following quantity is called as e-folding is important:
\begin{eqnarray}
&&N=\int_{t_i=\texttt{initial time}}^{t_f=\texttt{final time}}H(t)dt=\ln{\frac{a(t_f)}{a(t_i)}}\\&&\nonumber
=\frac{1}{q-1}\ln{\Big[1+(q-1)H_0 t\Big]}|_{t_i}^{t_f}\equiv ln_{q}(H_0 t)|_{t_i}^{t_f}.
\end{eqnarray}
Here $ln_{q}(z)$ is the inverse function of the $e_{q}(z)$.

\section{Single scalar $q$-inflaton}\label{scalar}
In the previous section we studied some cosmological consequences of a FLRW Universe when time evolution is determined by the $q$ - exponential form. One of the most important cases which it deserves to be investigated separately, is inflation. An era of early Universe which the time evolution is determined by de Sitter. We shall study our scenario of q-de Sitter in a simple form. \par
The most simple and applicable inflationary scenario is composed of a single slow varying scalar field. The typical action of such models is given by the following (For a review see \cite{Bassett:2005xm}):

\be
\label{S1}
S = \int d^4 x \sqrt{-g} \left( \frac{R}{2\kappa^2}
 - \frac{1}{2}\partial_\mu \phi \partial^\mu \phi - V(\phi) \right).
\ee
Here $V(\phi)$ denotes scalar potential. We write field equations in the form of effective quantities of a perfect fluid as the following:
\beqa
\rho=\frac12\dot{\phi}^2+V(\phi)\,,
~~~P=\frac12\dot{\phi}^2-V(\phi)\,,
\label{rhoeq}
\eeqa
where $\{P,\rho\}$ are pressure and energy density of inflaton respectly. Effective FLRW equations are written as follows:
\beqa
\label{Heq}
& &H^2 = \frac{8\pi}{3m_{\rm pl}^2}
\left[\frac12 \dot{\phi}^2+V(\phi) \right]\,, \\
& &\ddot{\phi}+3H\dot{\phi}+V_\phi(\phi)=0\,,
\label{phieq}
\eeqa
here
$V_{\phi} \equiv {\rm d}V/{\rm d}\phi$. \par
To proceed inflation, we require that $\dot{\phi}^2V(\phi)$. Classically it is equivalent that inflaton should behave like a tunneling particle, since in the inflationary era, the inflaton potential energy dominates over the kinetic energy.  Consequently we need a flat (almost flat) potential for inflation. A way to address this flat potential is to define a pair of
 slow-roll conditions as the following:
\beqa
\dot{\phi}^2/2 \ll V(\phi),\ \ |\ddot{\phi}| \ll 3H|\dot{\phi}|
\eeqa
We mention here that if we rewrite slow-roll conditions in terms of a dual $f (R) $ gravity, which is always available for a single scalar field, it leads to a second order corrected Einstein-Hilbert action which it nearly remains of the one proposed by $ f (R) =R+R^2/(6M^2) $ \cite{Sta80} .
In slow-roll approximation effective FLRW
Eqs.~(\ref{Heq}) and (\ref{phieq}) are given by the following:
\beqa
\label{eq1}
& & H^2 \simeq \frac{8\pi V(\phi)}{3m_{\rm pl}^2}\,, \\
& & 3H\dot{\phi}\simeq -V_\phi(\phi)\,.
\label{eq2}
\eeqa
Following literature, a pair of   slow-roll parameters $\epsilon$, $\eta$ and $\xi$ are written by the following :
\be
\label{S2}
\epsilon \equiv \frac{1}{2\kappa^2} \left( \frac{V'(\phi)}{V(\phi)} \right)^2\, ,\quad
\eta \equiv \frac{1}{\kappa^2} \frac{V''(\phi)}{V(\phi)}\, , \quad
\xi^2 \equiv \frac{1}{\kappa^4} \frac{V'(\phi) V'''(\phi)}{V(\phi)^2}\, .
\ee
Here $V' (\phi) \equiv \partial V(\phi)/\partial \phi$. Perturbations of early times are needed to produce matter. One effective quantitative parameter is
the tensor-to-scalar ratio which  is defined as
\be
\label{S3}
R = 16 \epsilon\, ,
\ee

and  later the spectral index $n_\mathrm{s}$ which it measures  the primordial curvature fluctuations and the corresponding running of the spectral index $\alpha_\mathrm{s}$ are given by:
\be
\label{SS1}
n_\mathrm{s} - 1 \sim - 6 \epsilon + 2 \eta\, , \quad
\alpha_\mathrm{s} \equiv \frac{d n_\mathrm{s}}{d \ln k}
\sim 16\epsilon \eta - 24 \epsilon^2 - 2 \xi^2
\, .
\ee
One can show that under slow-roll conditions, always we've $\epsilon \ll 1, \ \ |\eta| \ll 1$ and $|\xi^2| \ll 1$.
Inflation must be ended when when $\epsilon,|\eta|$ and $|\xi|^2\sim 1$, if this does not happen we shall introduce a reheating mechanism in our system. We review the basic problems of inflation and possible solutions to it.
To solve the flatness problem, it is needed that the relative density of inflation $\Omega_f$ to be $|\Omega_f-1|~\lsim~10^{-60}$ . It should be exactly right after the end of inflation.
Meanwhile the ratio $|\Omega-1|$ between the initial and final
phase of slow-roll inflation is given by
\beqa
\frac{|\Omega_f-1|}{|\Omega_i-1|} \simeq \left(\frac{a_i}{a_f}
\right)^2=e^{-2N_i}\,,
\label{2_3}
\eeqa
 We need to satisfy $|\Omega_i-1|\sim 1$ when  $N~\gsim~60$ to solve the flatness
problem.
\par
An exact inflationary solution of equations (\ref{eq1},\ref{eq2})  for $q$-de Sitter scale factor (\ref{a(t)})  is given by:
\beqa
V(\phi)=V_0exp\Big[\mp 2\gamma \phi\Big],
\label{v}
\eeqa
where
\beqa
V_0=\frac{3m_{pl}^2 H_0^2}{8\pi},\ \ \gamma=\frac{\sqrt{4\pi (q-1)}}{m_{pl}}\label{V}.
\eeqa
Such exponential potential form proposed in literature \cite{Lucchin and Matarrese,Yokoyam and Maeda,Khoury et al}. For this exponential potential, the set of low-roll parameters read:
\be
\label{S2}
\epsilon \equiv \frac{2\gamma^2}{\kappa^2} ,\quad
\eta \equiv 2\epsilon , \quad
\xi\equiv \eta .
\ee
For our model, the tensor-to-scalar ratio is obtained as
\be
\label{S3}
R = \frac{32\gamma^2}{\kappa^2}
\ee

and the spectral index and the associated running of the spectral index  are
\be
\label{SS1}
n_\mathrm{s} - 1 \sim -2\epsilon , \quad
\alpha_\mathrm{s} \equiv 0 .
\ee

We can easily verify that the above slow-roll approximations
are valid when $\gamma \ll \frac{\kappa}{\sqrt{2}}$.  The  upper bound
for this parameter is obtained by using WMAP9 and BICEP2 observational
data, $R < 0.36$ \cite{R} . For our model it gives us $\gamma < 0.1060660172 \kappa$. So our model is compatible with WMAP9 and BICEP2. In modified gravity also recently some models which are compatible with BICEP2 have been investigated \cite{Sebastiani:2013eqa}\par
To end the inflation, we should have $0<q(=\frac{3}{4})<1$. We deduce that the inflation ends when the evolution of the scale factor is $a (t) =e_ {3/4} (t) $. With this scale factor these is no specific $\phi_{end}$. So, we are not able to find the ending value of inflaton. In the next section we will propose another mechanism to solve this problem.

\section{Warm inflation}\label{warm}

Our $q$-inflation  in the context of single scalar field model (\ref{S1}), leads to the exponential potential (\ref{v}). This form of potential cannot describe  a complete inflationary theory. The slow-roll parameter $\epsilon$ (\ref{S2}) is constant, so the slow-roll era of inflation never ends and extra formalism is required to stop it. Exponential potential can not also fit well  the observational data \cite{1-w}. Inflaton field interacting with thermalised radiation may be solved this problem. This model is named "warm inflation," where dissipative effect is important during warm inflation and radiation production occurs with the inflationary expansion. Idea of warm inflation is phenomenologically explained by adding a friction term in the inflaton equation of motion \cite{2-w}.
\be
\label{w1}
\ddot{\phi}+(3H+\Gamma)\dot{\phi}+V_{,\phi}=0
\ee
$H(t)$ is the expansion rate and $\Gamma$ is damping term. We introduce an important parameter
\be
\label{w2}
r=\frac{\Gamma}{3H}
\ee
which is the relative strength of the thermal damping compared to the expansion damping. This parameter is much bigger  than one for warm inflation
and much lower than one for cool inflation.
Total energy density and pressure of warm inflation are modified
\begin{eqnarray}
\label{w3}
&&\rho_T=\rho_{\phi}+\rho_{\gamma}=\frac{1}{2}\dot{\phi}^2+V(\phi)+\rho_{\gamma}\\&&
\nonumber
P=\frac{1}{2}\dot{\phi}^2-V(\phi)
\end{eqnarray}
where $\rho_{\gamma}$ is energy density of radiation.
When the early universe is in homogeneous expansion the zero curvature Friedman equation is given by
\be
\label{w4}
3H^2=8\pi G(\rho_T)=8\pi G(\frac{1}{2}\dot{\phi}^2+V(\phi)+\rho_{\gamma})
\ee
Eqs.(\ref{w1}) and (\ref{w2}) are important for our future calculations.
Energy-momentum conservation
\be
\label{w5}
\dot{\rho}_T+3H(\rho_T+P)=0
\ee
is phenomenologically considered by two equations \cite{3-w}
\begin{eqnarray}
\label{w6}
&&\dot{\rho}_{\phi}+3H(\rho_{\phi}+P)=-\Gamma\dot{\phi}^2\\&&
\nonumber
\dot{\rho}_{\gamma}+4H(\rho_{\gamma})=\Gamma\dot{\phi}^2
\end{eqnarray}
where $\Gamma\dot{\phi}^2,$ is friction term. The energy-momentum conservation of scalar field (first segment of the above equations) is equivalent to Eq.(\ref{w1}).
In slow-roll limit, where the highest order in time derivative  is negligible, the above equations are simplified
\begin{eqnarray}
\label{w7}
&&\dot{\phi}=-\frac{V_{,\phi}}{3H(1+r)}\\&&
\nonumber
\rho_{\gamma}=\frac{3}{4}r\dot{\phi}^2\\&&
\nonumber
3H^2=8\pi GV
\end{eqnarray}
Note that $\rho_{\gamma}$ is the same order as two time derivatives so in Friedman equation we neglect it. Slow-roll approximation is governed by using slow-roll parameters. Slow-roll parameters of warm inflation model in high dissipative regime are given by \cite{4-w}
\begin{eqnarray}
\label{w8}
&&\epsilon=-\frac{1}{H}\frac{d}{dt}\ln H=\frac{1}{16\pi G r}(\frac{V_{,\phi}}{V})^2~~~~~~~~\\&&
\nonumber
\eta=-\frac{1}{2H}\frac{d}{dt}\ln(\dot{H}H\Gamma)=\frac{1}{8\pi G r}(\frac{V_{,\phi\phi}}{V})\\&&
\nonumber
\beta=-\frac{1}{H}\frac{d}{dt}(\ln \Gamma)=\frac{1}{8\pi G r}(\frac{\Gamma_{,\phi}V_{,\phi}}{\Gamma V})~~~~
\end{eqnarray}
Perturbation parameters of the warm inflation model are obtained in Ref.{\cite{4-w}}. Using these parameters, we can constrain the q-inflation model with the observational data\cite{WMAP}. Power-spectra, tensor to scalar ratio and spectral index of warm inflation are given by
\begin{eqnarray}
\label{w9}
&&P=(\frac{\pi}{4})^{\frac{1}{2}}\frac{H^{\frac{5}{2}}\Gamma^{\frac{1}{2}}T}{\dot{\phi}^2}~~~~~~~~~~~\\&&
\nonumber
R=\frac{32G}{\Gamma^{\frac{1}{2}}\pi^{\frac{3}{2}}T}\frac{\dot{\phi}^2}{H^{\frac{1}{2}}}~~~~~~~~~~~~~\\&&
\nonumber
n_s-1=-\frac{9}{4}\epsilon+\frac{3}{2}\eta-\frac{9}{4}\beta
\end{eqnarray}
Warm inflation models are studied in two important cases: 1)$\Gamma=\Gamma_0$ and 2)$\Gamma=\Gamma(\phi)=\Gamma_1V(\phi)$ \cite{5-w}. We will study q-inflation model in the context of warm inflation in these two cases.
\subsection{$\Gamma=\Gamma_0$}
Number of e-folds of q-inflation may be found using Eq.(\ref{q})
\begin{eqnarray}
\label{w10}
N=\int_{t_i}^{t_f} Hdt=\ln(a_f)-\ln(a_i)
\end{eqnarray}
where $t_i$ is the initial time of slow-roll inflation where $\epsilon=1$. Note that, we assume the system evolves in high dissipative regime i.e. $\Gamma\gg3H,$ and $\Gamma$ as a constant parameter. Using Eqs.(\ref{q}) and (\ref{w7}), we obtain the inflaton field in term of time.
\begin{eqnarray}
\label{w11}
\phi(t)=-(\frac{3H_0}{\pi G\Gamma_0(q-1)})^{\frac{1}{2}}\frac{1}{(1+(q-1)H_0 t)^{\frac{1}{2}}}
\end{eqnarray}
Potential in term of scalar field has the following form
\begin{eqnarray}
\label{w12}
V(\phi)=\frac{3H_0^2}{8\pi G\alpha_0^2}\phi^4
\end{eqnarray}
where ~$\alpha_0=\frac{3H_0}{\Gamma_0\pi G(1-q)}$~.
The slow-roll parameters of the model in this case are given by
\begin{eqnarray}
\label{w13}
&&\epsilon=\frac{(1-q)^2}{\alpha_0^q}\phi^{2q}\\&&
\nonumber
\eta=\frac{(1-q)(3-2q)}{2\alpha_0^q}\phi^{2q}\\&&
\nonumber
\beta=0
\end{eqnarray}
Slow-roll inflation ($\ddot{a}>0$) occurs where $\epsilon<1,$ from above equation this condition only satisfied when $\phi^{2q}<\frac{2\alpha^q}{(1-q)^2}$.
The number of efolds is presented by
\begin{eqnarray}
\label{w14}
N=\ln a|_{\phi=\phi}-\ln a|_{\phi_i}=\frac{1}{1-q}(\ln \phi^2_i-\ln\phi^2)
\end{eqnarray}
We assume the slow-roll inflation begins at $\epsilon=1,$ so the scalar field at this time becomes
\begin{eqnarray}
\label{w15}
\phi_i=\frac{\alpha_0^{\frac{1}{2}}}{(1-q)^{\frac{1}{q}}}
\end{eqnarray}
From two above equations, inflaton in term of number of e-folds is presented by
\begin{eqnarray}
\label{w16}
\phi^2(N)=\frac{\alpha_0}{(1-q)^{\frac{2}{q}}}\exp(-(1-q)N)
\end{eqnarray}
Using Eq.(\ref{w9}), we can find the perturbation parameters of the model. These parameters may be related to observational data \cite{WMAP}. Using Eq.(\ref{w7}) and (\ref{w9}) we present the power-spectrum in term of scalar field.
\begin{eqnarray}
\label{w17}
P_R=\frac{3H_0^3(1-q)}{4\pi G\alpha_0^3}\phi^6
\end{eqnarray}
Spectral index is given by
\begin{eqnarray}
\label{w18}
&& n_s(\phi)=1+\frac{q(1-q)}{2\alpha_0^q}\phi^{2q}\\&&
\nonumber
n_s(N)=1+\frac{q}{2(1-q)}\exp(-(1-q)N)
\end{eqnarray}

\begin{figure} [ht]
\centering
\includegraphics[width=10cm,height=10cm,keepaspectratio]{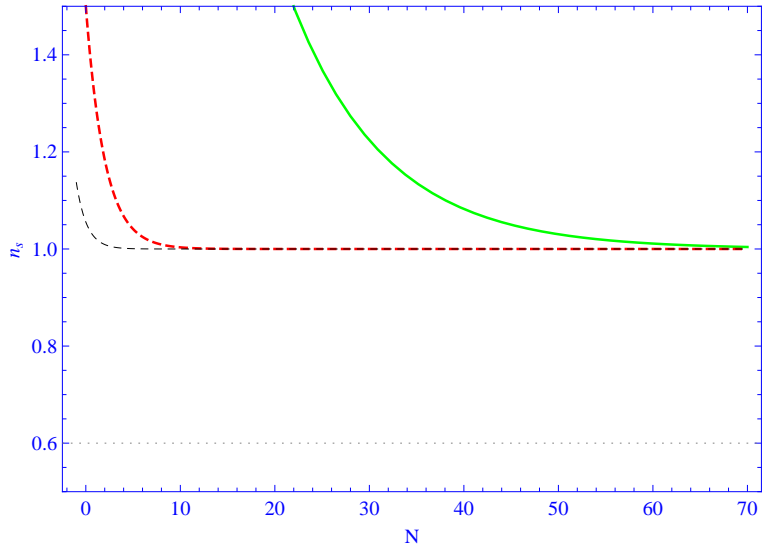}
\caption{ Spectral index $n_s$ in term of number of e-folds, from left to right: $q=\frac{1}{10},~~q=\frac{1}{2},~~ q=\frac{9}{10}$.}
\label{the-label-for-cross-referencing}
\end{figure}
\begin{figure} [ht]
\centering
\includegraphics[width=10cm,height=10cm,keepaspectratio]{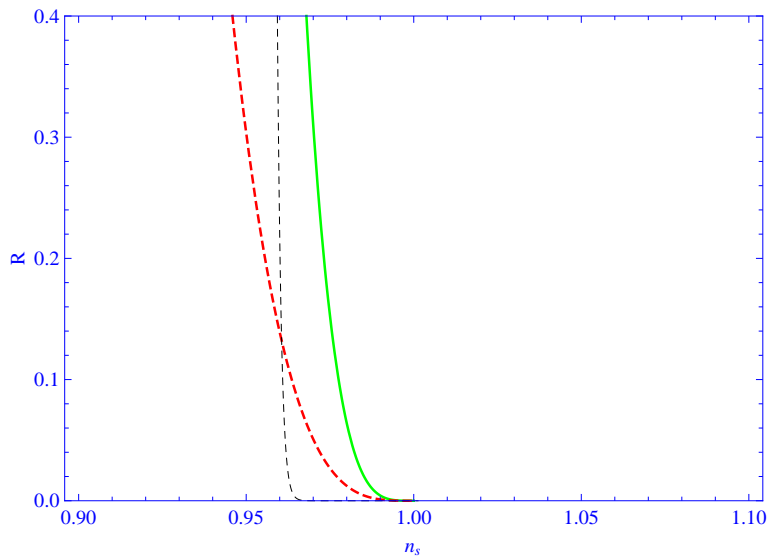}
\caption{Tensor-scalar ratio  $R$ in term of spectral index $n_s$.}
\label{the-label-for-cross-referencing}
\end{figure}
FIG.(1) shows the scale invariant spectrum, (Harrison-Zeldovich spectrum, i.e. $n_s=1$) may be approximately presented by ($q=\frac{9}{10},~~N=70$).
Tensor-scalar ratio in term of scalar field and spectral index are given by
\begin{eqnarray}
\label{w19}
&& R(\phi)=\frac{24G H_0^{\frac{5}{2}}(1-q)}{\Gamma_0^{\frac{1}{2}}\pi^{\frac{3}{2}}T\alpha_0^{\frac{5}{2}}}\phi^7\\&&
\nonumber
R(n_s)=\frac{24G H_0^{\frac{5}{2}}(1-q)}{\Gamma_0^{\frac{1}{2}}\pi^{\frac{3}{2}}T\alpha_0^{\frac{5}{2}}}(\frac{2\alpha^q}{q(1-q)})^{\frac{7}{2q}}(1-n_s)^{\frac{7}{2q}}
\end{eqnarray}

In Fig.(2), three trajectories in the $n_s-R$ plane are shown. There is a range of values of $R$ and $n_s$ which is compatible with the BICEP2 data where $q=\frac{9}{10}$.
\subsection{$\Gamma=\Gamma_1V(\phi)$}
Dissipative coefficient may be considered as a function of inflaton field $\phi$ \cite{5-w}. Using Eqs.(\ref{q}) and (\ref{w7}), we present the inflaton field in this case.
\begin{eqnarray}
\label{w20}
\phi=\alpha_1(1+(1-q)H_0t)^{\frac{1}{2}}
\end{eqnarray}
where $\alpha_1=(\frac{8}{\Gamma_1(1-q)H_0})^{\frac{1}{2}}$. Potential of the model in term of scalar field has the following form
\begin{eqnarray}
\label{w21}
V(\phi)=\frac{3H_0^2\alpha_1^4}{8\pi G}\frac{1}{\phi^4}
\end{eqnarray}
We also present the slow-roll parameters in this case
\begin{eqnarray}
\label{w22}
&&\epsilon=\frac{(1-q)^2\alpha_1^{2q}}{\phi^{2q}}\\&&
\beta=2(1-q)\\&&
\nonumber
\eta=\frac{\alpha_1^{2q}(1-q)(3-2q)}{2\phi^{2q}}+\frac{\beta}{2}
\end{eqnarray}
Using above equation we can find the initial scalar field where $\epsilon=1$
\begin{eqnarray}
\label{w23}
\phi_i=\alpha_1(1-q)^{\frac{1}{q}}
\end{eqnarray}
The scalar field in term of number of e-folds is presented from Eqs.(\ref{w10}) and (\ref{w23})
\begin{eqnarray}
\label{w24}
\phi^2(N)=\alpha_1^2(1-q)^{\frac{2}{q}}\exp(-(1-q)N)
\end{eqnarray}
Perturbation parameters of warm q-inflation, in this case, could help us to constrain the model with observational data. Power-spectrum is presented by using the above equations
\begin{eqnarray}
\label{w25}
P_{R}=\frac{3T^2\alpha^5}{32G(1-q)^2H_0^2}\frac{1}{\phi^{\frac{5}{2}}}
\end{eqnarray}
Spectral index in term of inflaton and number of e-folds has the following form
\begin{eqnarray}
\label{w26}
&&n_s-1=\frac{3(1-q)q\alpha_1^{2q}}{4\phi^{2q}}+3(1-q)\\&&
\nonumber
~~~~~~~\simeq\frac{3}{4}q\exp(-q(1-q)N)
\end{eqnarray}
\begin{figure} [ht]
\centering
\includegraphics[width=10cm,height=10cm,keepaspectratio]{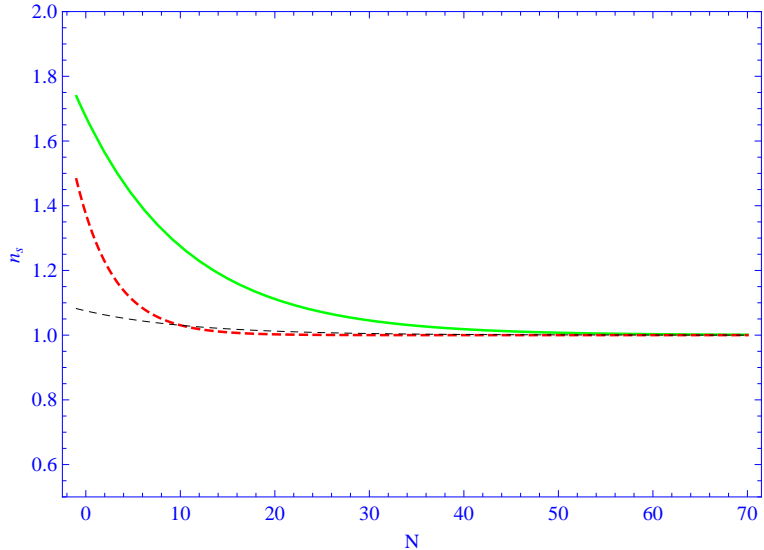}
\caption{Spectral index $n_s$ in term of number of e-folds $N$.}
\label{the-label-for-cross-referencing}
\end{figure}
\begin{figure} [ht]
\centering
\includegraphics[width=10cm,height=10cm,keepaspectratio]{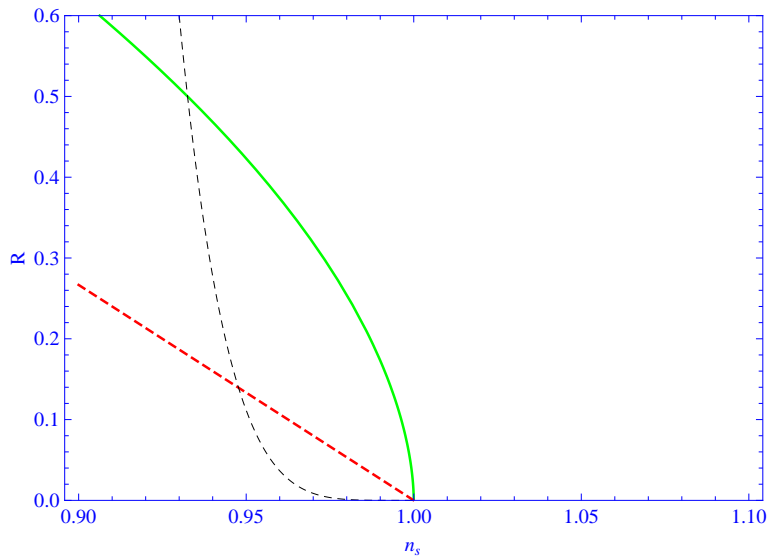}
\caption{Tensor-scalar ratio  $R$ in term in term of pectral index $n_s$.}
\label{the-label-for-cross-referencing}
\end{figure}
In Fig.(3), the dependence of spectral index on the number of e-folds of inflation is shown (for $q=\frac{1}{10}$, $q=\frac{1}{2}$ and $\frac{9}{10}$ cases). It is  observed that small values of number of e-folds are assured for small  values of $q$ parameter. The scale-invariant spectrum (Harrison-Zeldovich spectrum, i.e. $n_s=1$)  may be presented by $(q,N)=(\frac{9}{10},70)$.
Tensor-scalar ratio in this case is given by
\begin{eqnarray}
\label{w27}
&&R=\frac{A}{\phi}\\&&
\nonumber
~~=A(\frac{4(1-n_s)}{3q(1-q)\alpha_1^{2q}})^{\frac{1}{2q}}
\end{eqnarray}
where $A=\frac{128\sqrt{2}G^{\frac{3}{2}}(1-q)\alpha_1}{\sqrt{3}(H_0\Gamma_1)^{\frac{3}{2}}\pi T}$.

In Fig.(4), three trajectories in the $n_s-R$ plane are shown. For $q=\frac{9}{10}$  a range of values of $R$ and $n_s$  is compatible with the BICEP2 data.

\section{conclusion}\label{conclusion}
In this article, we have proposed a novel inflationary mechanism driven by a family of non extensive de Sitter scale factors. Such scale factors belongs to a wide family of statistical functions, is called as q-generalized functions. Our model -$q$ de Sitter- is a new class of cosmological solutions driven by an intermediate scale factor. Our model of inflation, $q$-de Sitter inflation opens a new bridge between non extensive systems and cosmology. We show that this scale factor gives rise to a unified scenario of dark energy and dark matter. 
The single inflaton scenario is investigated. A constant ratio $R$ is obtained. It has been proven that the model is compatible with observational data. But such single inflaton scenario with exponential potential has no ending point. To end inflation, we also study warm inflation with two forms of damping factors. In each case, we show that inflationary model has exact solutions for inflaton and interaction potential. We show that for specific values of non extensively parameter $q$, a range of values of $R$ and $n_s$ is compatible with the BICEP2 data. This is the first work in which the non extensive statistical assumptions are described.

\begin{acknowledgments} We gratefully acknowledge M. Sami for useful comments.
\end{acknowledgments}


\clearpage
\appendix


\begin{thebibliography}{99}

\bibitem{Sta80}
 A. A. Starobinsky,JETP Lett. 30, 719(1979);
A. A. Starobinsky, Phys. Lett. 91B, 99(1980);
A. A. Starobinsky, Phys. Lett. 117B, 175(1982);
A. A. Starobinsky, JETP Lett. 55, 489(1992);
A. A. Starobinsky, S. Tsujikawa, and J. Yokoyama, Nucl.
Phys. B 610, 383(2001).


\bibitem{oldinf}
D.~Kazanas,
Astrophys.\ J.\  {\bf 241} L59 (1980);
K.~Sato, Mon.\ Not.\ R.\ Astron.\ Soc. {\bf 195}, 467 (1981);
Phys.\ Lett.\ {\bf 99B}, 66 (1981);
A.~H.~Guth,
Phys.\ Rev.\ D {\bf 23}, 347 (1981).

\bibitem{WMAP}
G.~Hinshaw {\it et al.}  [WMAP Collaboration],
  Astrophys.\ J.\ Suppl.\  {\bf 208}, 19 (2013)
  [arXiv:1212.5226 [astro-ph.CO]];
  P.~A.~R.~Ade {\it et al.}  [Planck Collaboration],
  arXiv:1303.5082 [astro-ph.CO]; P.~A.~R.~Ade {\it et al.}  [Planck Collaboration],
  arXiv:1303.5076 [astro-ph.CO].

\bibitem{LLbook}
A.~R.~Liddle and D.~H. ~Lyth, {\em Cosmological inflation and
large-scale structure}, Cambridge University Press (2000).
\bibitem{COBE}
G.~F.~Smoot {\it et al.},
Astrophys.\ J.\  {\bf 396}, L1 (1992).

\bibitem{WMAP1}
D.~N.~Spergel {\it et al.}  [WMAP Collaboration],
Astrophys.\ J.\ Suppl.\  {\bf 148}, 175 (2003).

\bibitem{Planck}
P.~A.~R.~Ade {\it et al.}  [Planck Collaboration],
arXiv:1303.5076 [astro-ph.CO].

\bibitem{oldper}
V.~F.~Mukhanov and G.~V.~Chibisov,
JETP Lett.\  {\bf 33}, 532 (1981);
A.~H.~Guth and S.~Y.~Pi,
Phys.\ Rev.\ Lett.\  {\bf 49} (1982) 1110;
S.~W.~Hawking,
Phys.\ Lett.\ B {\bf 115}, 295 (1982);
A.~A.~Starobinsky,
Phys.\ Lett.\ B {\bf 117} (1982) 175;
J.~M.~Bardeen, P.~J.~Steinhardt and M.~S.~Turner,
Phys.\ Rev.\ D {\bf 28}, 679 (1983).
\bibitem{mog}
 S. Nojiri and S. D. Odintsov, Phys. Rept. 505, 59 (2011) [arXiv:1011.0544 [gr-qc]];K. Bamba, S. Nojiri and S. D. Odintsov, arXiv:1302.4831 [gr-qc];K. Bamba and S. D. Odintsov, arXiv:1402.7114 [hep-th]; S. Nojiri and S. D. Odintsov, AIP Conf. Proc. 1115, 212 (2009) [arXiv:0810.1557
[hep-th]]; S. Capozziello and M. De Laurentis, Phys. Rept. 509, 167 (2011) [arXiv:1108.6266
[gr-qc]];S. Capozziello, L. Consiglio, M. De Laurentis, G. De Rosa and C. Di Donato,
arXiv:1110.5026 [astro-ph.CO].

\bibitem{Riess}
A.~G.~Riess {\it et al.}  [Supernova Search Team Collaboration],
Astron.\ J.\  {\bf 116}, 1009 (1998)
[astro-ph/9805201].

\bibitem{Perlmutter}
S.~Perlmutter {\it et al.}  [Supernova Cosmology Project Collaboration],
Astrophys.\ J.\  {\bf 517}, 565 (1999)
[astro-ph/9812133].

\bibitem{BAO}
D.~J.~Eisenstein {\it et al.}  [SDSS Collaboration],
Astrophys.\ J.\  {\bf 633}, 560 (2005)
[astro-ph/0501171].

\bibitem{Weinberg}
S.~Weinberg,
Rev.\ Mod.\ Phys.\  {\bf 61}, 1 (1989).

\bibitem{CST}
E.~J.~Copeland, M.~Sami and S.~Tsujikawa,
Int.\ J.\ Mod.\ Phys.\ D {\bf 15}, 1753 (2006)
[hep-th/0603057].

\bibitem{quinpapers}
Y.~Fujii, Phys.\ Rev.\ D {\bf 26}, 2580 (1982);
L.~H.~Ford,
Phys.\ Rev.\ D {\bf 35}, 2339 (1987);
C.~Wetterich, Nucl. \ Phys \ B. {\bf 302}, 668 (1988);
T.~Chiba, N.~Sugiyama and T.~Nakamura,
Mon.\ Not.\ Roy.\ Astron.\ Soc.\  {\bf 289}, L5 (1997)
[astro-ph/9704199];
N.~Majd and D.~Momeni,
  Int.\ J.\ Mod.\ Phys.\ E {\bf 20}, 113 (2011)
  [arXiv:0903.2020 [gr-qc]];
D.~Momeni and A.~Azadi,
  Astrophys.\ Space Sci.\  {\bf 317}, 231 (2008)
  [arXiv:0903.0081 [gr-qc]];
M.~Khurshudyan, N.~S.~Mazhari, D.~Momeni, R.~Myrzakulov and M.~Raza,
  arXiv:1403.0081 [gr-qc];
P.~G.~Ferreira and M.~Joyce,
Phys.\ Rev.\ Lett.\  {\bf 79}, 4740 (1997)
[astro-ph/9707286];
R.~R.~Caldwell, R.~Dave and P.~J.~Steinhardt,
Phys.\ Rev.\ Lett.\  {\bf 80}, 1582 (1998)
[astro-ph/9708069];
M.~Jamil, D.~Momeni and R.~Myrzakulov,
  Chin.\ Phys.\ Lett.\  {\bf 29} (2012) 109801
  [arXiv:1209.2916 [physics.gen-ph]];

M.~Jamil, D.~Momeni and R.~Myrzakulov,
  Eur.\ Phys.\ J.\ C {\bf 72}, 2267 (2012)
  [arXiv:1212.6017 [gr-qc]];
T.~Chiba, N.~Sugiyama and T.~Nakamura,
Mon.\ Not.\ Roy.\ Astron.\ Soc.\  {\bf 301}, 72 (1998)
[astro-ph/9806332];
I.~Zlatev, L.~-M.~Wang and P.~J.~Steinhardt,
Phys.\ Rev.\ Lett.\  {\bf 82}, 896 (1999)
[astro-ph/9807002].


\bibitem{3} A. Berera, Warm inflation, Phys. Rev. Lett. 75 (1995) 3218 [astro-ph/9509049];
A.~Berera,
  Phys.\ Rev.\ D {\bf 55}, 3346 (1997)
  [hep-ph/9612239].
\bibitem{Tsujikawa:2014mba}
  S.~Tsujikawa,
  arXiv:1404.2684 [gr-qc].


\bibitem{WMAP9}
G.~Hinshaw {\it et al.}  [WMAP Collaboration],
Astrophys.\ J.\ Suppl.\  {\bf 208}, 19 (2013)
[arXiv:1212.5226 [astro-ph.CO]].

\bibitem{CDT}
T.~Chiba, A.~De Felice and S.~Tsujikawa,
Phys.\ Rev.\ D {\bf 87}, 083505 (2013)
[arXiv:1210.3859 [astro-ph.CO]].

\bibitem{moreview}
T.~P.~Sotiriou and V.~Faraoni,
Rev.\ Mod.\ Phys.\  {\bf 82}, 451 (2010)
[arXiv:0805.1726 [gr-qc]];
A.~De Felice and S.~Tsujikawa,
Living Rev.\ Rel.\  {\bf 13}, 3 (2010)
[arXiv:1002.4928 [gr-qc]];
 R.~Myrzakulov, L.~Sebastiani and S.~Zerbini,
 Int.\ J.\ Mod.\ Phys.\ D {\bf 22}, 1330017 (2013)
 [arXiv:1302.4646 [gr-qc]];
 G.~Cognola, O.~Gorbunova, L.~Sebastiani and S.~Zerbini,
 Phys.\ Rev.\ D {\bf 84}, 023515 (2011)
 [arXiv:1104.2814 [gr-qc]];
M.~J.~S.~Houndjo, D.~Momeni and R.~Myrzakulov,
  Int.\ J.\ Mod.\ Phys.\ D {\bf 21}, 1250093 (2012)
  [arXiv:1206.3938 [physics.gen-ph]];
M.~Jamil, S.~Ali, D.~Momeni and R.~Myrzakulov,
  Eur.\ Phys.\ J.\ C {\bf 72}, 1998 (2012)
  [arXiv:1201.0895 [physics.gen-ph]];
D.~Momeni and M.~R.~Setare,
  Mod.\ Phys.\ Lett.\ A {\bf 26}, 2889 (2011)
  [arXiv:1106.0431 [physics.gen-ph]];
M.~R.~Setare and D.~Momeni,
  Int.\ J.\ Theor.\ Phys.\  {\bf 50}, 106 (2011)
  [arXiv:1001.3767 [physics.gen-ph]];
D.~Momeni,
  Int.\ J.\ Theor.\ Phys.\  {\bf 50}, 1493 (2011)
  [arXiv:0910.0594 [gr-qc]];
M.~Jamil, D.~Momeni and R.~Myrzakulov,
  Gen.\ Rel.\ Grav.\  {\bf 45}, 263 (2013)
  [arXiv:1211.3740 [physics.gen-ph]];
 G.~Cognola, L.~Sebastiani and S.~Zerbini,
 arXiv:1301.3031 [gr-qc];
S.~Tsujikawa,
Lect.\ Notes Phys.\  {\bf 800}, 99 (2010)
[arXiv:1101.0191 [gr-qc]];
T.~Clifton, P.~G.~Ferreira, A.~Padilla and C.~Skordis,
Phys.\ Rept.\  {\bf 513}, 1 (2012)
[arXiv:1106.2476 [astro-ph.CO]].

\bibitem{5-w}  M.~R.~Setare and V.~Kamali,
  JHEP {\bf 1303}, 066 (2013)
  [arXiv:1302.0493 [hep-th]];
  Phys.\ Rev.\ D {\bf 87}, 083524 (2013)
  [arXiv:1305.0740 [hep-th]];
  JCAP {\bf 1208}, 034 (2012)
  [arXiv:1210.0742 [hep-th]];
  arXiv:1407.2604 [gr-qc].


\bibitem{Jamil:2013nca}
  M.~Jamil, D.~Momeni and R.~Myrzakulov,
  arXiv:1309.3269 [gr-qc].
\bibitem{slow-roll}
A. D. Linde, Phys. Lett. B 108, 389 (1982); A. Albrecht and P. J. Steinhardt, Phys. Rev.
Lett. 48, 1220 (1982); A. D. Linde, Phys. Lett. B 129, 177 (1983).
\bibitem{k-inflation}
C. Armendariz-Picon, T. Damour and V. F. Mukhanov, Phys. Lett. B 458, 209 (1999)
[arXiv:hep-th/9904075].
\bibitem{extended}
D. La and P. J. Steinhardt, Phys. Rev. Lett. 62, 376 (1989) [Erratum-ibid. 62, 1066 (1989)].
\bibitem{Higgs}
C. Germani, A. Kehagias, Phys. Rev. Lett. 105, 011302 (2010). [arXiv:1003.2635 [hep- ph]]; C. Germani, A. Kehagias, JCAP 1005, 019 (2010). [arXiv:1003.4285 [astro-ph.CO]]; C. Germani, A. Kehagias, Phys. Rev. Lett. 106, 161302 (2011). [arXiv:1012.0853 [hep-ph]


\bibitem{Tsallis}
C. Tsallis, J. Stat. Phys. 52, 479 (1988);
See http://tsallis.cat.cbpf.br/biblio.htm for a regularly updated bibliography on the
subject.
\bibitem{Exton}
Exton, H, "q-Hypergeometric Functions and Applications", New York: Halstead Press, Chichester: Ellis Horwood, 1983, ISBN 0853124914, ISBN 0470274530, ISBN 978-0470274538 (1983); Ferhan M. Atici \& Paul W. Eloe (2007)," Fractional q-Calculus on a time scale", Journal of Nonlinear
Mathematical Physics, 14:3, 341-352, DOI: 10.2991/jnmp.2007.14.3.4.


\bibitem{Tsallis (2009)}
C. Tsallis, "Introduction to nonextensive statistical mechanics : approaching a complex world" , New York: Springer. ISBN 978-0-387-85358-1(2009).




\bibitem{Martin:2013nzq}
  J.~Martin, C.~Ringeval, R.~Trotta and V.~Vennin,
  JCAP {\bf 1403}, 039 (2014)
  [arXiv:1312.3529 [astro-ph.CO]].




\bibitem{Lucchin and Matarrese}F. Lucchin , and S. Matarrese, Phys. Rev. D 32, 1316 (1985).

\bibitem{Yokoyam and Maeda}
J. Yokoyama , and K. i. Maeda,  Phys. Lett. B 207, 31 (1988).

\bibitem{Khoury et al}
J. Khoury , B. A. Ovrut, P. J. Steinhardt, and N. Turok,
Phys. Rev. D 64, 123522 (2001).

\bibitem{R}
 G.~Hinshaw {\it et al.}  [WMAP Collaboration],
  Astrophys.\ J.\ Suppl.\  {\bf 208}, 19 (2013)
  [arXiv:1212.5226 [astro-ph.CO]];
  P.~A.~R.~Ade {\it et al.}  [Planck Collaboration],
  arXiv:1303.5082 [astro-ph.CO]; P.~A.~R.~Ade {\it et al.}  [Planck Collaboration],
  arXiv:1303.5076 [astro-ph.CO]; P. A. R. Ade et al. [BICEP2 Collaboration], arXiv:1403.3985 [astro-ph.CO]; P. A. R. Ade et al.
[BICEP2 Collaboration], arXiv:1403.4302 [astro-ph.CO].

\bibitem{Sebastiani:2013eqa}
 L.~Sebastiani, G.~Cognola, R.~Myrzakulov, S.~D.~Odintsov and S.~Zerbini,
 Phys.\ Rev.\ D {\bf 89}, 023518 (2014)
 [arXiv:1311.0744 [gr-qc]];
 K.~Bamba, R.~Myrzakulov, S.~D.~Odintsov and L.~Sebastiani,
result,''
 Phys.\ Rev.\ D {\bf 90}, 043505 (2014)
 [arXiv:1403.6649 [hep-th]]. 
\bibitem{Bassett:2005xm}
  B.~A.~Bassett, S.~Tsujikawa and D.~Wands,
  Rev.\ Mod.\ Phys.\  {\bf 78}, 537 (2006)
  [astro-ph/0507632].

\bibitem{1-w} I.~Dalianis and F.~Farakos,
  arXiv:1405.7684 [hep-th].
\bibitem{2-w} A. Hosoya and M. aki Sakagami, Phys. Rev. D 29, 2228 (1984).
\bibitem{3-w} A. Berera, Phys. Rev. Lett. 75, 3218 (1995).
\bibitem{4-w} L.~M.~H.~Hall, I.~G.~Moss and A.~Berera,
  Phys.\ Rev.\ D {\bf 69}, 083525 (2004)
  [astro-ph/0305015].



\end{thebibliography}
\end{document}